\documentclass[12pt,preprint]{aastex}

\newcommand{\chandra}{{\em Chandra}}
\newcommand{\xmm}{{\em XMM-Newton}}
\newcommand{\suzaku}{{\em Suzaku}}
\newcommand{\swift}{{\em Swift}}
\newcommand{\fuse}{{\em FUSE}}
\newcommand{\euve}{{\em EUVE}}
\newcommand{\hst}{{\em HST}}
\newcommand{\spitzer}{{\em Spitzer}}
\newcommand{\vla}{{\em VLA}}
\newcommand{\gemini}{{\em Gemini}}

\newcommand{\src}{NGC 4051}

\newcommand{\nh}{N\ensuremath{_{\rm H} }}	
\newcommand{\tin}{T\ensuremath{_{\rm in} }}	
\newcommand{\rin}{r\ensuremath{_{\rm in} }}	
\newcommand{\nj}{N\ensuremath{_{\rm j} }}	
\newcommand{\te}{T\ensuremath{_{\rm e} }}	
\newcommand{\zacc}{z\ensuremath{_{\rm acc} }}	
\newcommand{\ro}{r\ensuremath{_{\rm 0} }}	
\newcommand{\ho}{h\ensuremath{_{\rm 0} }}	
\newcommand{\rg}{\ensuremath{r_{\rm g} }}	
\newcommand{\msun}{M\ensuremath{_\odot}}	
\newcommand{\about}{\ensuremath{\sim}}		
\newcommand{\bh}{black hole}			
\newcommand{\ledd}{L\ensuremath{_{\rm Edd} }}	
\newcommand{\mbh}{M\ensuremath{_{\rm BH} }}	

\newcommand{\valinclin}{50\arcdeg}
\newcommand{\valnj}{1.7\ensuremath{\times10^{-2} }} 
\newcommand{\valro}{4}                              
\newcommand{\valhratio}{1}
\newcommand{\valte}{4\ensuremath{\times10^{10} }}   
\newcommand{\valzacc}{5}                            
\newcommand{\valpe}{2.7}
\newcommand{\valrin}{130}                           
\newcommand{\valtin}{2.8\ensuremath{\times10^4}}    
\newcommand{\valLeddmodel}{\ensuremath{10^{-2} }} 

\shorttitle{Broadband SED of \src}
\shortauthors{Maitra et al.}
\begin{document}
\title{%
A Jet Model for the Broadband Spectrum of the Seyfert-1 Galaxy NGC 4051
}

\author{
 Dipankar Maitra\altaffilmark{1}, 
 Jon M. Miller\altaffilmark{1},
 Sera Markoff\altaffilmark{2}, and
 Ashley King\altaffilmark{1}
}
\affil{Department of Astronomy, University of Michigan,
    Ann Arbor, MI 48109, USA}
\affil{Astronomical Institute ``Anton Pannekoek'', 
    University of Amsterdam, P.O. Box 94249, 1090 GE Amsterdam, 
    The Netherlands}
\email{dmaitra@umich.edu}

\begin{abstract}   
Recent radio VLBI observations of the $\sim$parsec-scale nuclear
region of the narrow line Seyfert 1 galaxy \src\ hint toward the
presence of outflowing plasma.  From available literature we have
collected high-quality, high-resolution broadband spectral energy
distribution data of the nuclear region of \src\ spanning from radio
through X-rays, to test whether the broadband SED can be explained
within the framework of a relativistically outflowing jet model.
We show that once the contribution from the host galaxy is taken
into account, the broadband emission from the active galactic nucleus
of \src\ can be well described by the jet model.  Contributions
from dust and ongoing star-formation in the nuclear region tend to
dominate the IR emission even at the highest resolutions.  In the
framework of the jet model, the correlated high variability of the
extreme ultraviolet and X-rays compared to other wavelengths suggests
that the emission at these wavelengths is optically thin synchrotron
originating in the particle acceleration site(s) in the jet very
close (few \rg=$G\mbh/c^2$) to the central supermassive black hole
of mass \mbh.  Our conclusions support the hypothesis that narrow
line Seyfert 1 galaxies (which \src\ is a member of) harbor a
``jetted'' outflow with properties similar to what has already been
seen in low-luminosity AGNs and stellar mass black holes in hard
X-ray state.
\end{abstract}


\section{Introduction} \label{s:intro} 
While it is generally agreed that the enormous energy output of
active galactic nuclei (AGN) as well as X-ray binaries (XRB) is due
to accretion of matter onto a compact object ($GM/[c^2R]\about 1$,
see, e.g., \citealt{Rees1984} for a review), the fate of the accreted
material is somewhat uncertain.  Most likely, a large fraction of
this accreted matter crosses the event horizon of the black hole
and is forever lost from the observer's view.  However, the presence
of relativistic outflows, often dubbed ``jets'', observed to emanate
from the very core of the AGNs and XRBs, show that some of the
accreted material is in fact ejected from the inner accretion flow.
Recent simulations show that such jets are likely Poynting flux
dominated, i.e., only a very small fraction of the accreted material
is ejected while most of the energy in the jets is carried by the
electromagnetic fields.  Launching of such relativistic outflows,
their composition, and their association with the accretion inflow
are currently topics of intense research.  Observations indicate
that nuclear radio-loudness of AGNs (parametrized by the radio-to-optical
luminosity ratio) tends to increase with decreasing Eddington-scaled
nuclear luminosity \citep{Ho2002, Sikora+2007}.  A similar situation
is seen in XRBs accreting at low mass accretion rates where the
radio loudness (in this case parametrized by radio to soft X-ray
flux ratio) tends to increase with decreasing X-ray luminosity
\citep{Fender+2003}.  The radio emission from a radio loud AGN is
generally attributed to synchrotron emission from a population of
nonthermal particles in the jets and the lobes. But the emission
mechanism in radio quiet AGNs such as Seyferts remains much less
well understood.  Recent VLBI observations reported by \citet{GP2009}
have detected radio emission from the nuclear region of four very
faint Seyfert galaxies (NGC 4051, NGC 4388, NGC 4501, and NGC 5033),
where high brightness temperature (T$_{\rm B}\sim 10^5$--$10^7$ K)
and (in some cases) resolved structure in the core favor a weak
jet/outflow origin for these sources.

Among these four faint Seyferts studied by \citet{GP2009}, \src,
one of the ``original'' Seyfert galaxies studied by \citet{Seyfert1943}
due to its strong nuclear emission lines, is a narrow line Seyfert
1 galaxy (NLS1; see \citealt{op1985} for definition of NLS1s) that
has been well studied in almost all wavelengths from radio to the
X-rays.  Located in the Ursa Major cluster of galaxies, \src\ is a
fairly nearby (z=$0.002336\pm0.000004$) spiral galaxy.  Based on
the latest reverberation mapping estimates, the AGN in the heart
of \src\ boasts a supermassive black hole (SMBH) of $(1.7\pm0.5)\times10^6$
\msun\ \citep{Denney+2009}. As mentioned above, recent VLBI
observations of \src\ at 1.6 and 5 GHz show a complex, sub-parsec-scale
compact core and other structure \citep{GP2009}.  Similar results
have also been obtained from recent \vla/{\em E}\vla\ observations
\citep{King+2011,Jones+2011}.  The high brightness temperature of
the compact core of \src\ ($>1.8\times10^5$ K) along with the
observed radio structures lead to the conclusion that the radio
emission maybe be nonthermal in origin, and likely originating from
an outflow/jet.

From Gemini Near-infrared Integral Field Spectrograph (NIFS) data
of \src, \citet{Riffel+2008} have found a strong H\,{\sc ii}
Br$\gamma$ emission at 2.1661 $\mu$m in the nuclear spectrum (see
their Fig.  1). Based on H$_{\rm 2}\lambda$2.1218$\mu$m/Br$\gamma$
line ratio they concluded that some contribution to the strong
Br$\gamma$ emission is from shocks or energy injection due to
interaction of the H$_{\rm 2}$ emitting gas with the jet.  While
the strong Br$\gamma$ line may be associated with interaction with
the jet, it is quite unlikely that the IR continuum emission is
dominated by the jet. Earlier high time resolution monitoring at
2.2$\mu$m continuum emission of \src\ by \citet{Hunt+1992} does not
show any rapid IR variability (relative to X-rays) which might be
expected if the dominant fraction of the observed IR emission is
an extension of the optically-thin X-ray emission (e.g.  as recently
found for the stellar black hole binary system XTE~J1550$-$564 by
\citealt{Russell+2010}).  Simultaneous optical, IR and X-ray
observations of \src\ obtained by \citet{Done+1990} also show little
optical/IR variability compared to the X-rays, suggesting different
origins for the X-ray continuum and optical/IR continuum emission.

Compact, steady jets are observed from XRBs when they are in {\em
hard} X-ray state, i.e., their X-ray spectrum is dominated by hard
power law emission and large r.m.s.\ variability.  In contrast,
such steady, compact jets are not seen when XRBs are in the {\em
soft} X-ray state characterized by predominantly thermal spectrum
from the accretion disc and low r.m.s. variability.  See, e.g.,
\citet{HomanBelloni2005,RM2006}, for extensive definitions and
review of X-ray states in XRBs, and \citet{Fender2006} for a review
of XRB jets.  It is still not clear which physical mechanism(s)
trigger transition between these states, as well as the associated
change in disk--jet coupling.  Such state transitions and
appearance/disappearance of radio jets are now regularly seen in
transient X-ray binary systems (where epochs of activity typically
span \about month or less), thanks to advances in instrumentation
and coordinated, quasi-simultaneous multiwavelength observations
over the past decade.  Such changes however cannot be seen in AGN
because the associated timescales can easily exceed the human
lifespan, and advances can only be made via studying ensemble
properties of AGN. Indeed, recent works show that there exists a
tight relationship connecting the radio luminosity, X-ray luminosity,
and black hole mass.  This relationship, often referred to as the
``fundamental plane'' of black hole accretion
\citep{Merloni+2003,Falcke+2004}, not only holds across the stellar
mass black holes in hard state and low-luminosity AGNs (LLAGN), but
also for supermassive black holes with direct, dynamical mass-measurements
(i.e. the $M-\sigma$ AGNs; see \citealt{Gultekin+2009}), and extends
to blazars as well (Plotkin et al. in prep.).

\citet{McHardy+2004} have argued that \src\ and the stellar mass
black hole Cygnus X--$1$ in soft state have very similar Fourier
power density spectra, and therefore \src\ may be a soft state AGN
accreting at \about30\% \ledd.  Based on the bending frequency of
the Fourier power spectrum of \src, these authors had deduced a
central black hole mass of $3^{+2}_{-1}\times10^5$\msun, which was
consistent with the best estimated reverberation mass of
$5^{+6}_{-3}\times10^5$\msun\ at that time.  However, the most
recent, revised reverberation mass estimate of
$1.73^{+0.55}_{-0.52}\times10^6$\msun\ by \citet{Denney+2009} is
\about5.7 times larger, consequently lowering the bolometric
luminosity to \about5\% \ledd.  Moreover, as discussed in detail
in their discussion section, the assumed X-ray to bolometric
conversion factor of 27 is quite likely an overestimate. On the
other hand, another commonly used empirical relationship to estimate
the bolometric luminosity of similar AGNs is $L_{\rm bol}\approx9\lambda
L_\lambda$ (5100 \AA) \citep{Kaspi+2000, Peterson+2004}. Given
\src's host galaxy starlight subtracted average nuclear flux density
at 5100 \AA\ is (4.5$\pm$0.4)$\times10^{-15}$ erg s$^{-1}$ cm$^{-2}$
\AA$^{-1}$ \citep{Denney+2009}, its bolometric luminosity is roughly
$\approx(1\%)(D/10\,{\rm Mpc})^2(\mbh/1.7\times10^{6}\msun)^{-1}$
\ledd. Both of the above estimates of the bolometric luminosity are
closer to Galactic black hole binaries in hard state than in soft
state.  It is also worth noting that Cygnus X-1 is actually never
in a ``classical'' soft state, but rather it is always somewhat
transitional (see, e.g., \citealt{Nowak+2005}, \citealt{Wilms+2006}).
Thus it is unclear whether \src\ is indeed a soft state AGN, or in
hard state, or perhaps in a transitory state between canonical hard
and soft states.  Can XRB state definitions be straightforwardly
carried over to AGNs?  Such are the questions which motivate the
current work.

Since jets are known to emit over a broad range of the electromagnetic
spectrum spanning from radio through X-rays, it is plausible that
at least some of the observed nuclear emission from \src\ might
also originate in the jet.  In this work we have collected high
quality, high-resolution nuclear SEDs of \src\ from the literature,
to study the broadband (radio--X-ray) SED and test whether the
observed nuclear emission can originate in a mildly relativistic
jet outflow.  We use the accretion disk+jet model developed by
\citet{Markoff+2005}, \citet{Maitra+2009a} to model the broadband
continuum.  The model has been successfully applied to explaining
broadband SEDs of several black hole X-ray binaries in hard state,
as well as that of the LLAGNs Sagittarius A*, M81* \citep[see,
e.g.,][and references therein]{Markoff2010}. These previous results
suggest that despite a difference of \about5--6 orders of magnitude
in mass, the accretion/ejection processes in stellar mass black
holes (\about 10\msun) and supermassive black holes in the nearby
low-luminosity AGNs (\about $10^{6-7}$\msun) appear quite similar
when comparing sources with similar Eddington scaled luminosity.

This is the first time we apply this jet model to a radio-quiet
Seyfert galaxy to test whether the broadband emission originates
in a collimated jet outflow.  The sources from which we obtained
the multi-wavelength SED data are described in \S\ref{s:data}, and
a brief summary of the model parameters that best match the data
is given in \S\ref{s:analysis}. Implications of the modeling results
and conclusions drawn from the study are discussed in \S\ref{s:conclusion}.

\section{The broadband SED of \src} \label{s:data} 
The nuclear region of \src\ has been observed with many different
ground-based as well as space-borne telescopes spanning from radio
through X-rays, at the highest spatial resolution available in each
of these wavelength ranges.  The high angular resolution is of
crucial importance for isolating the central, nuclear emission from
that of the surrounding galaxy. This is especially important in
optical and IR wavelengths where contributions from non-AGN components
such as the host galaxy's stellar population, ongoing star formation
near the nuclear region, and dust can easily dominate the observed
emission if the photometric aperture is not small.

\begin{figure} 
\centering
 \includegraphics[height=0.95\textwidth, angle=-90]{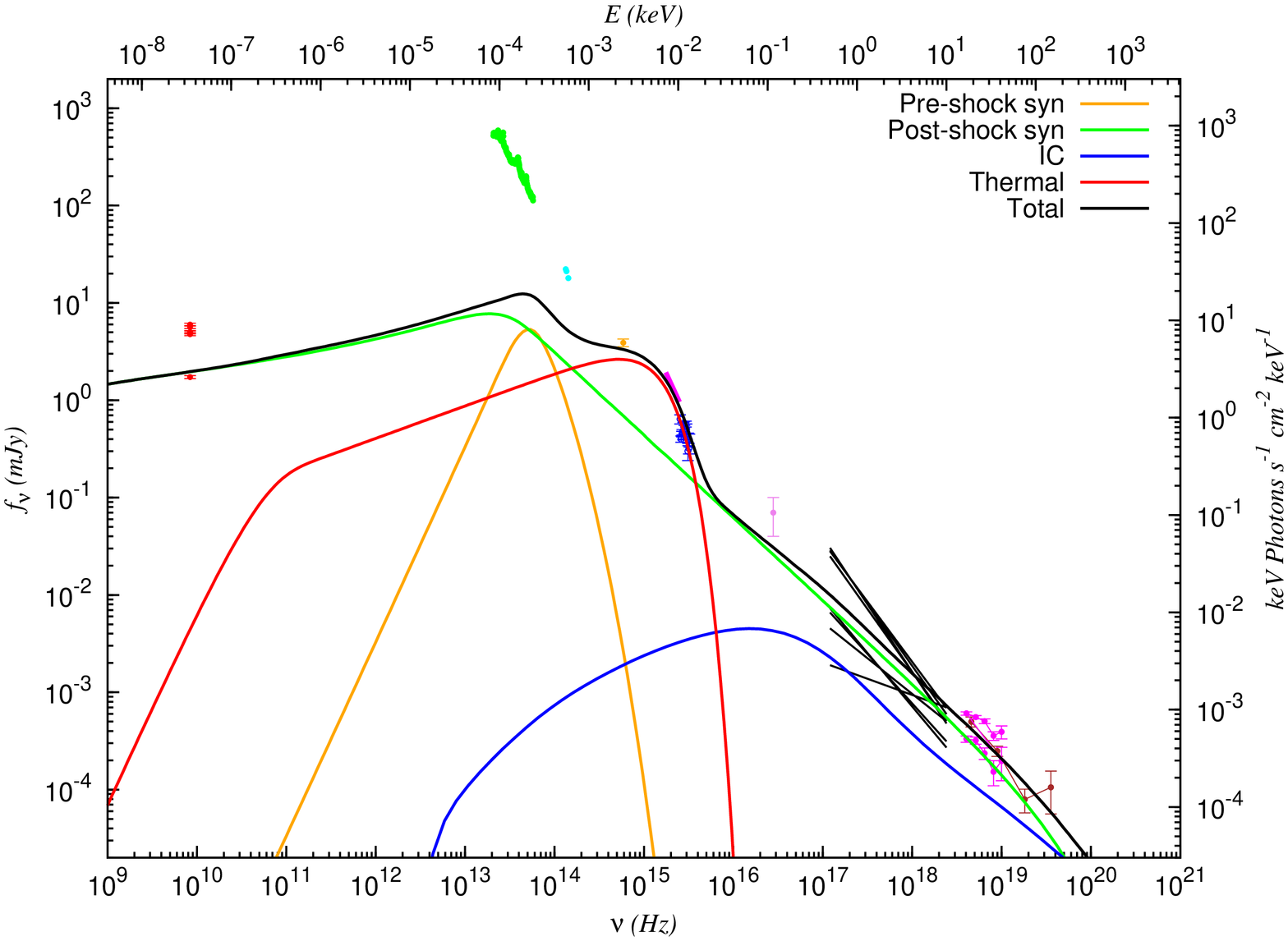} 
   \caption{Broadband SED of \src. 
   \chandra/ACIS (solid black lines) and \vla\ 8.4 GHz (red points) data taken 
   from \citet{King+2011}.
   \suzaku/PIN data (magenta points connected by lines) from \citet{Miller+2010}.
   \swift/BAT data (brown points connected by line) from publicly available 
   NASA archives.
   \euve\ data (violet) from \citet{Uttley+2000},
   \fuse\ data (blue points) are from \citet{Kaspi+2004},
   \hst/STIS fluxes (magenta line) from \citet{Collinge+2001}, and
   5100\AA\ continuum data point (orange) from \citet{Denney+2009}.
   The \gemini/NIFS data \citep{Riffel+2008} and the \spitzer/IRS data 
   \citet{Sani+2010} are shown in cyan and green respectively.
   The SED was corrected for Galactic absorption assuming a column density
   of E(B--V)=0.013 and \nh=1.3$\times$10$^{20}$ cm$^{-2}$ in the direction 
   of \src\ \citep{Elvis+1989}, and assuming the extinction law of 
   \citet{Cardelli+1989}.  The fits to high-resolution \chandra\ spectra
   discussed in \citet{King+2011} suggest that intrinsic
   absorption is negligible. The thick solid lines show the various model 
   components as well as the total model predicted SED.
   }
 \label{f:sed} 
\end{figure}

{\subsection{X-ray}}  
The X-ray flux from \src\ is known to be highly variable on a
timescale of days \citep[see, e.g.,][]{McHardy+2004,Ponti+2006,Breedt+2010,
Vaughan+2011}.  Like most NLS1s, the 2--10 keV X-ray continuum SED
of \src\ is well modeled by a power law.  A narrow iron line is
seen near 6.4 keV, and the presence of curvature at higher energies
hint towards presence of reflection \citep{Ponti+2006, Miller+2010}.
Similar to the X-ray spectra of many other NLSs, there exists an
excess of soft photons in addition to the power law at $<$2 keV
\citep[see, e.g.,][also see \S\ref{s:conclusion}]{Boller+1996}.  In
this work we are interested in the power-law continuum only and
take the power-law photon index ($\Gamma$) and normalization from
a recent simultaneous radio (\vla/{\em EVLA})--X-ray (\chandra)
campaign of \src\ published by \citet{King+2011}. We chose this set
of X-ray observations because they were performed simultaneously
with the radio observations at 8.4 GHz.  The solid black lines in
Fig.~\ref{f:sed} show the power-law continuum measured from this
\chandra\ data set.
As discussed in \S\ref{s:conclusion}, the hard X-ray ($>$10 keV)
variability may be smaller than the soft (2--10 keV) X-ray variability.
Therefore we have also added two {\em Suzaku}/PIN spectra of \src\
from the data presented by \citet{Miller+2010}.
While \src\ has not been detected in the {\em Fermi}/LAT data so
far \citep[0.1--100 GeV;][]{Abdo+2010}, it has been detected in the
time averaged {\em Swift}/BAT (14--195 keV) data.  The BAT
spectrum\footnote{Obtained from
http://heasarc.gsfc.nasa.gov/docs/swift/results/bs9mon/bs\_ind.php?number=85}
is also included in Fig.~\ref{f:sed}.
Fits to the high-resolution \chandra\ HETGS spectra of \src\ (King
et al., in prep.) suggest that any absorption intrinsic to the
source is negligible, therefore to correct for absorption we use a
column density of $\nh=1.3\times10^{20}$ cm$^{-2}$ in the direction
of \src\ \citep{Elvis+1989}.

{\subsection {UV}} 
The {\em Extreme Ultraviolet Explorer (EUVE)} data are from
\citet{Uttley+2000}, and the corresponding errorbar represents the
EUV variability range (also see Fig. 7 of \citealt{Kaspi+2004}).
\citet{Uttley+2000} noted that the EUV variability is comparable
to the X-ray variability, and that the EUV and X-rays were simultaneous
to within 1 ksec, which suggests a common origin for both EUV and
soft X-ray emission.

The far-ultraviolet data in the wavelength range of 900--1180 \AA\
are taken from the {\em Far Ultraviolet Spectroscopic Explorer}
(\fuse) observations presented by \citet{Kaspi+2004}.  Even though
the three \fuse\ observations were separated by as much as $\sim$
one year, the analysis by \citet{Kaspi+2004} suggests little or no
($<$10\%) variation in the \fuse\ spectra.  As discussed above, we
are interested in the continuum emission only, and therefore use
the average flux-density measurements in wavelength ranges that are
free from emission and absorption as presented in Table 2 of
\citet{Kaspi+2004}.

We have also included \hst/STIS data of \src\ published by
\citet{Collinge+2001}.  The binned STIS spectrum presented in their
work shows that the continuum flux is roughly constant at
$\sim2\times10^{-14}$ erg~cm$^{-2}$~s$^{-1}$~\AA$^{-1}$ over the
observed wavelength range of 1200--1700 \AA.  These \hst\ observations
were made simultaneously with \chandra, and even though the \chandra\
light curve showed large variations in the X-ray flux (see, e.g,
Fig. 2 of \citealt{Collinge+2001}), no variability could be detected
in the STIS data.  The lack of UV variability strongly contrasts
with the high X-ray variability and suggests that the UV and X-rays
are most likely coming from different emission mechanisms.

{\subsection {Optical}}  
Contribution of non-AGN components, e.g. light from stars as well
as dust, starts becoming important in optical and IR, and extreme
caution must be taken not only to use highest resolution available
to observe the nuclear region, but also to model and subtract the
non-AGN contributions within the aperture.  Readily available optical
and IR fluxes listed, e.g. in the NED, are therefore not usable for
this work since the apertures listed in NED are typically large and
there is no easy way to estimate the non-AGN contribution.  Instead
we use the results obtained by \citet{Denney+2009}, who presented
new, optical spectroscopic and photometric observations of \src\
from a densely sampled reverberation mapping campaign, using data
from five different observatories (1.3m McGraw-Hill telescope at
MDM Observatory, 0.7 and 2.6m telescopes of the Crimean Astrophysical
Observatory (CrAO), 2m MAGNUM telescope at the Haleakala Observatories,
and the 0.4m telescope of the University of Nebraska(UNebr)).  They
then subtract the host galaxy starlight contribution to the continuum
flux following the prescription by \citet{Bentz+2009}.  The estimated
AGN continuum flux at 5100 \AA\ for \src\ is $(4.5\pm0.4)\times10^{-15}$
erg~s$^{-1}$~cm$^{-2}$~\AA$^{-1}$ after subtracting host galaxy
starlight contribution.  The host galaxy starlight contribution is
about twice that of the AGN.  However the optical flux is much less
variable than the X-ray and EUV.  The excess variance in the 5100
\AA\ measurements of \citet{Denney+2009} is 0.09. A recent analysis
of last twelve years of X-ray and optical variability in \src\ by
\citet{Breedt+2010} also concluded that the variability of the AGN
in optical wavelengths (where the fractional variability, F$_{\rm
var}$, is $\sim$12--26\%) is much smaller than the the variability
in X-rays (F$_{\rm var}\sim49$\%).

{\subsection {Infrared}} 
For IR we used the \spitzer\ mid-infrared (MIR; 5--15 $\mu$m)
spectrum of \src\ published by \citet{Sani+2010}, and also adaptive
optics Integral Field Unit spectroscopic data reported by
\citet{Riffel+2008} which was obtained with the NIFS mounted on the
Gemini North telescope.  Three representative continuum fluxes at
2.1, 2.2, and 2.25$\mu$m from the nuclear (central
0.1\arcsec$\times$0.1\arcsec) region obtained by the NIFS are shown
by the cyan points in Fig.~\ref{f:sed}.

The \spitzer\ spectrum (also see Fig. 2 of \citealt{Sani+2010})
shows strong polycyclic aromatic hydrocarbon (PAH) features due to
ongoing star formation activity. According to \citet{Sani+2010},
the MIR continuum may also be primarily due to thermal emission by
AGN-heated dust.  Therefore contribution of direct AGN emission in
MIR appears to be small.  The slit width of the IR spectrograph is
3.6 arcsec, which at the distance of \src\ implies a linear scale
of $\sim$175 (D/10 Mpc) parsec. The \spitzer/IRS slit width is a
factor of 18 larger than that of the \hst/STIS slit width (0.2
arcsec) and the former therefore includes significantly larger
circumnuclear regions.  The \spitzer\ data therefore do not put
strong constraints on our broadband modeling.  However they put
upper limits on the direct (accretion/ejection origin) AGN emission
in the MIR wavelengths. The Gemini NIFS nuclear spectrum with an
angular coverage of $0.1\arcsec\times0.1\arcsec$, is expected to
contain a significantly larger fraction of AGN-to-non-AGN (starburst,
dust) emission compared to the IRS.

{\subsection {Radio}} 
The 8.4 GHz radio data are taken from \citet{King+2011}, where the
authors present simultaneous radio and X-ray monitoring observations
of \src.  During this campaign \src\ was observed 6 times between
2008 December 31 and 2009 July 31 simultaneously using both \vla/{\em
EVLA} GHz and \chandra. Details of the radio data analysis are
presented in \citet{King+2011}.

\bigskip
The UV, optical and IR data were dereddened assuming E(B--V)=0.013
magnitudes and using the extinction law of \citet{Cardelli+1989}.
We note that no cosmological K-correction was made to the observed
continuum fluxes. Given the small redshift of \src\ (z=0.0023), any
K-correction is negligible and uncertainties in absolute flux
calibrations are much larger than this correction.
Table~\ref{tab:observations} gives the wavelength range and observation
dates as well as references to the published works we have used to
construct the broadband SED.

\section{Modeling the SED} \label{s:analysis} 
The continuum jet emission in the outflow-dominated model we have
used here is based on previous works by \citet{Maitra+2009a} and
\citet{Markoff+2005}.  We refer the reader to these works for the
details of the model.  The main parameters that determine the
properties of the jet in this model are the input jet power (\nj),
electron temperature of the relativistic thermal plasma entering
at the jet base (\te), the ratio of magnetic to particle energy
density (a.k.a. the equipartition factor $k$), physical dimensions
of the jet base (assumed to be cylindrical with radius \ro\ and
height \ho), and the location of the point on the jet (\zacc) beyond
which a fraction of the leptons are accelerated to a power-law
energy distribution ($dN/dE \about E^{-p}$). \nj, parametrized in
terms of the Eddington luminosity, determines the power initially
input into the particles and magnetic field at the base of the jets.
Beyond the jet-base, we assume that the jet plasma (a) accelerates
mildly due to longitudinal pressure gradient with its bulk Lorentz
factor asymptotically reaching a value of 2--3 far downstream along
the jet, (b) expands laterally with its initial sound speed, and
(c) cools due to adiabatic expansion.  Once the initial conditions
are specified, then the bulk speed of the jet plasma, magnetic
field, and particle density are computed by solving the adiabatic,
relativistic Euler equation \citep{Falcke1996}.  We note that in
AGN, a combination of higher powers, confinement and perhaps stronger
magnetic field pressures may give higher bulk Lorentz factors.

The accretion disk is modeled as a multi-color blackbody with radial
temperature profile T$\propto$R$^{-0.75}$ \citep{ss1973, Mitsuda+1984},
and is parametrized by \tin, \rin, the temperature and radius at
the inner edge of the disc.  The thermal photons from the accretion
disc are also included in the photon field of the jet for Compton
scattering, although due to Doppler redshifting in the jet frame,
this external Compton emission is usually much smaller than the
synchrotron self-Compton.

\begin{figure} 
\centering
 \includegraphics[height=0.43\textwidth, angle=-90]{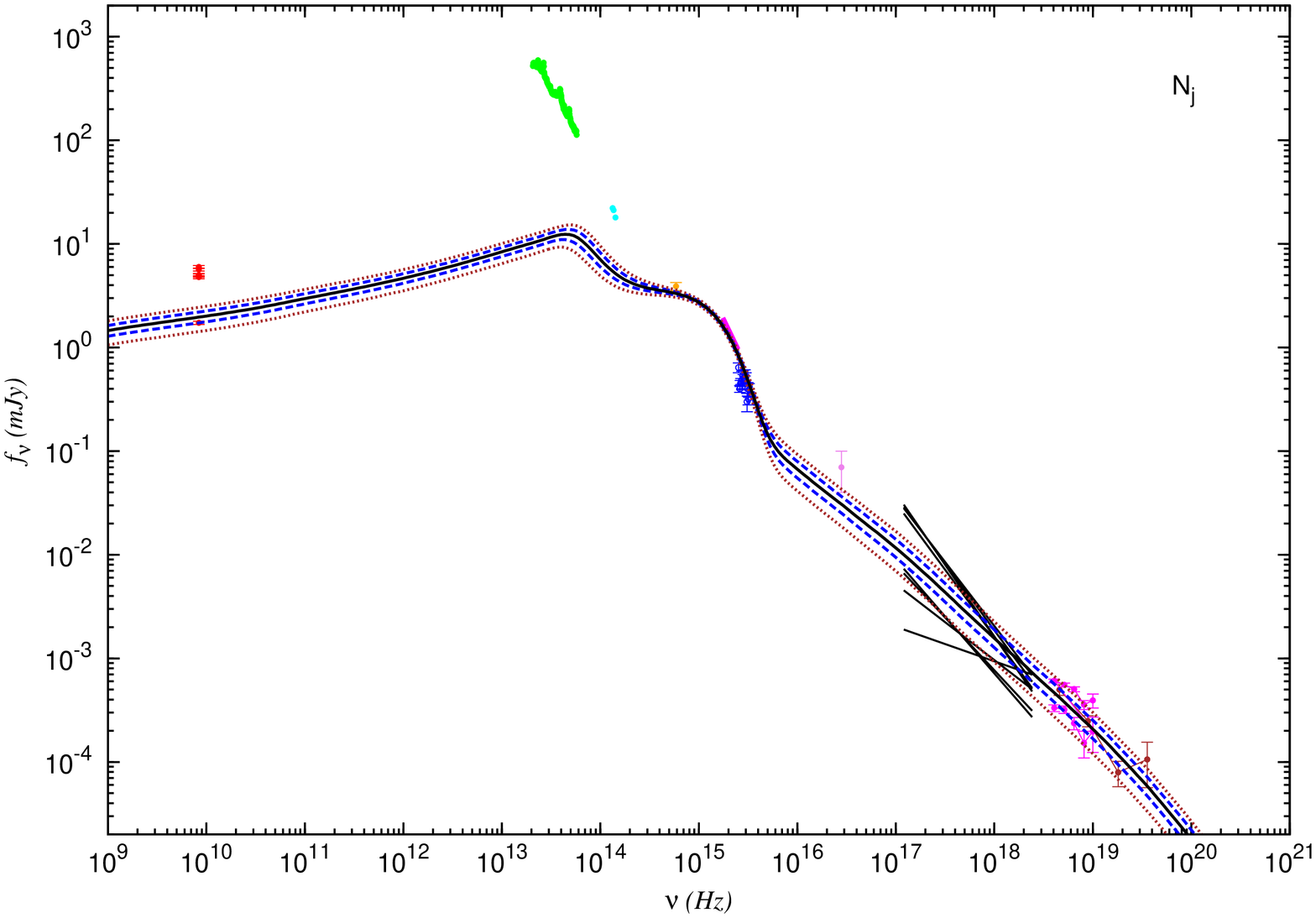} 
 \includegraphics[height=0.43\textwidth, angle=-90]{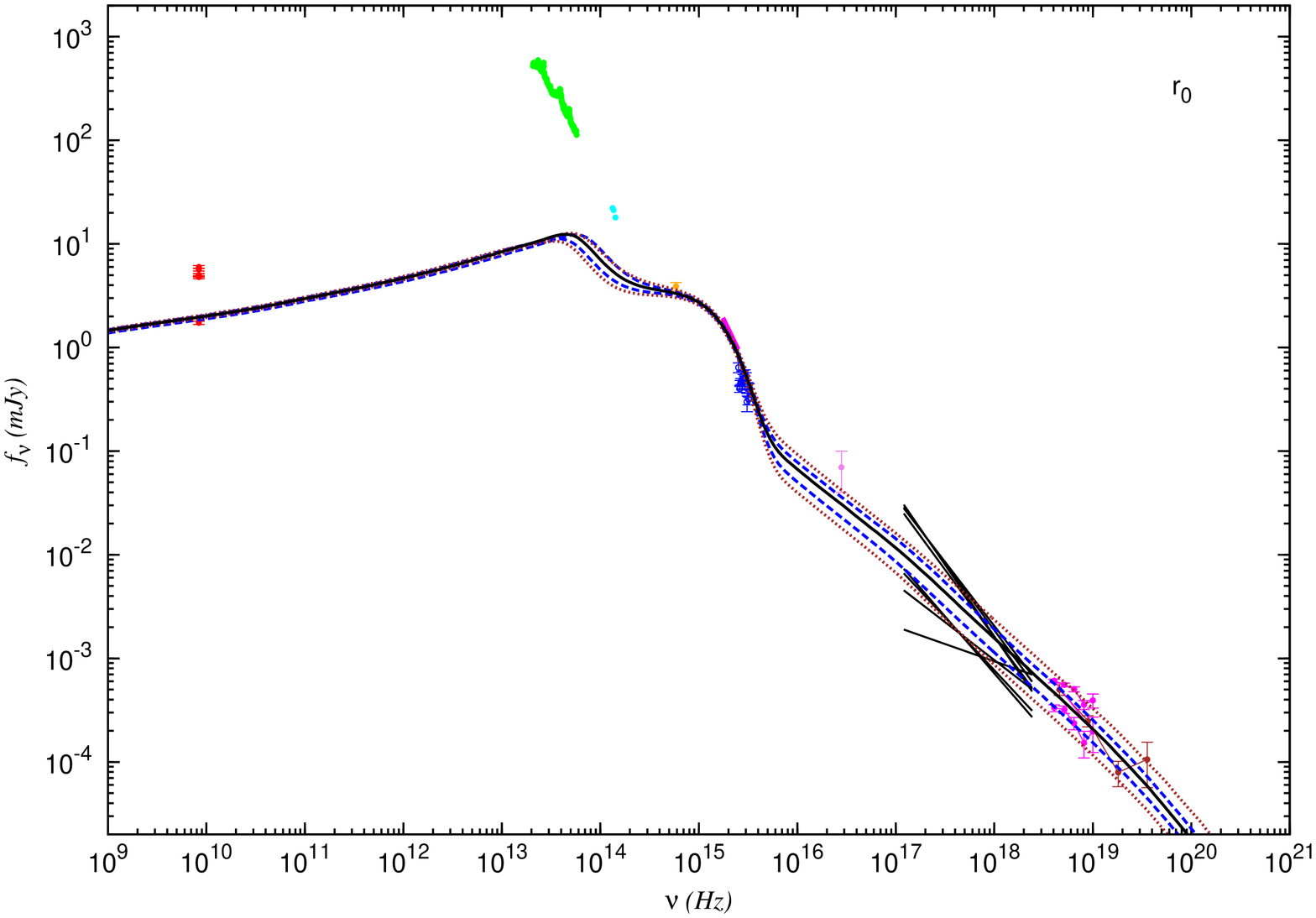} 
 \includegraphics[height=0.43\textwidth, angle=-90]{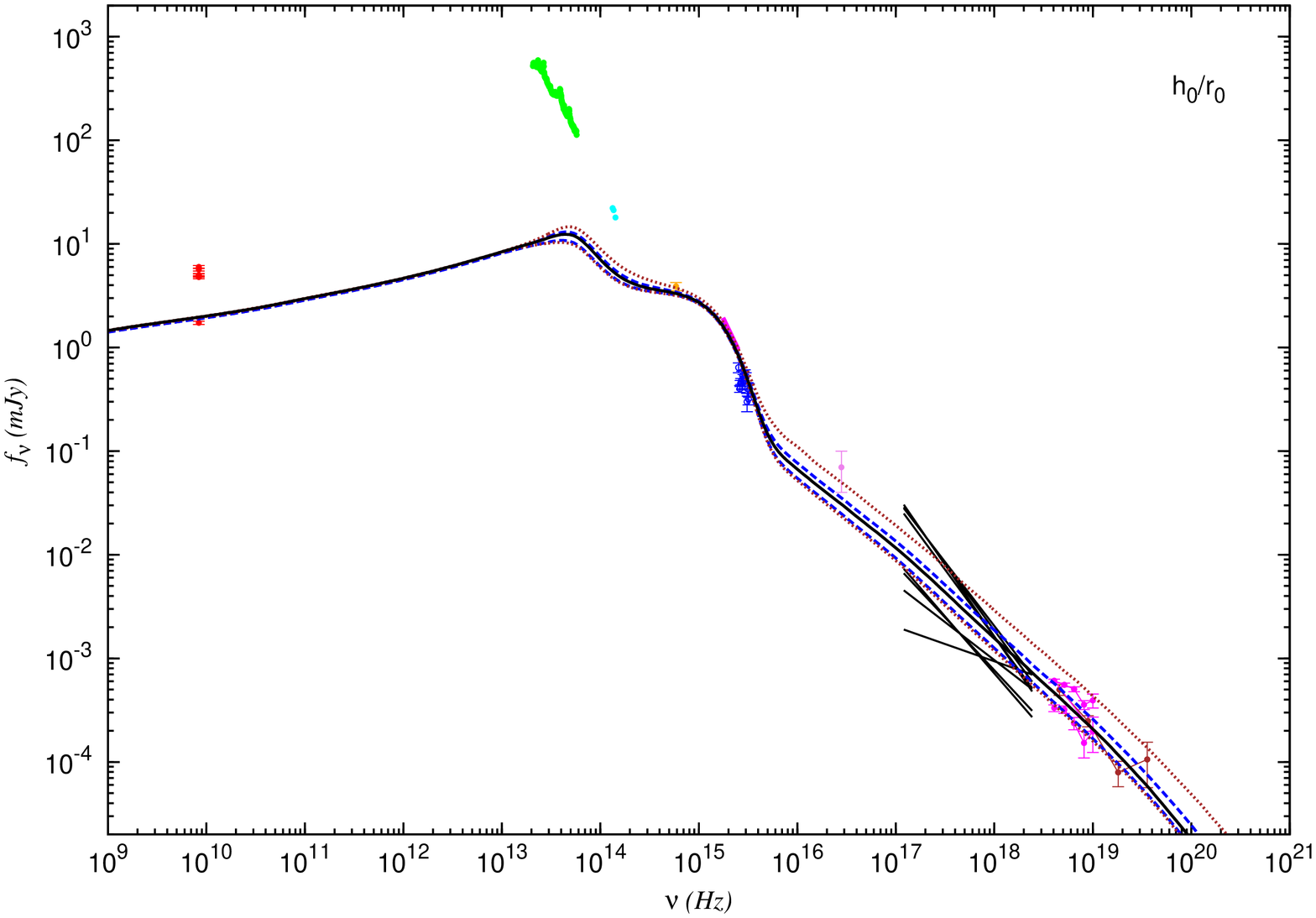} 
 \includegraphics[height=0.43\textwidth, angle=-90]{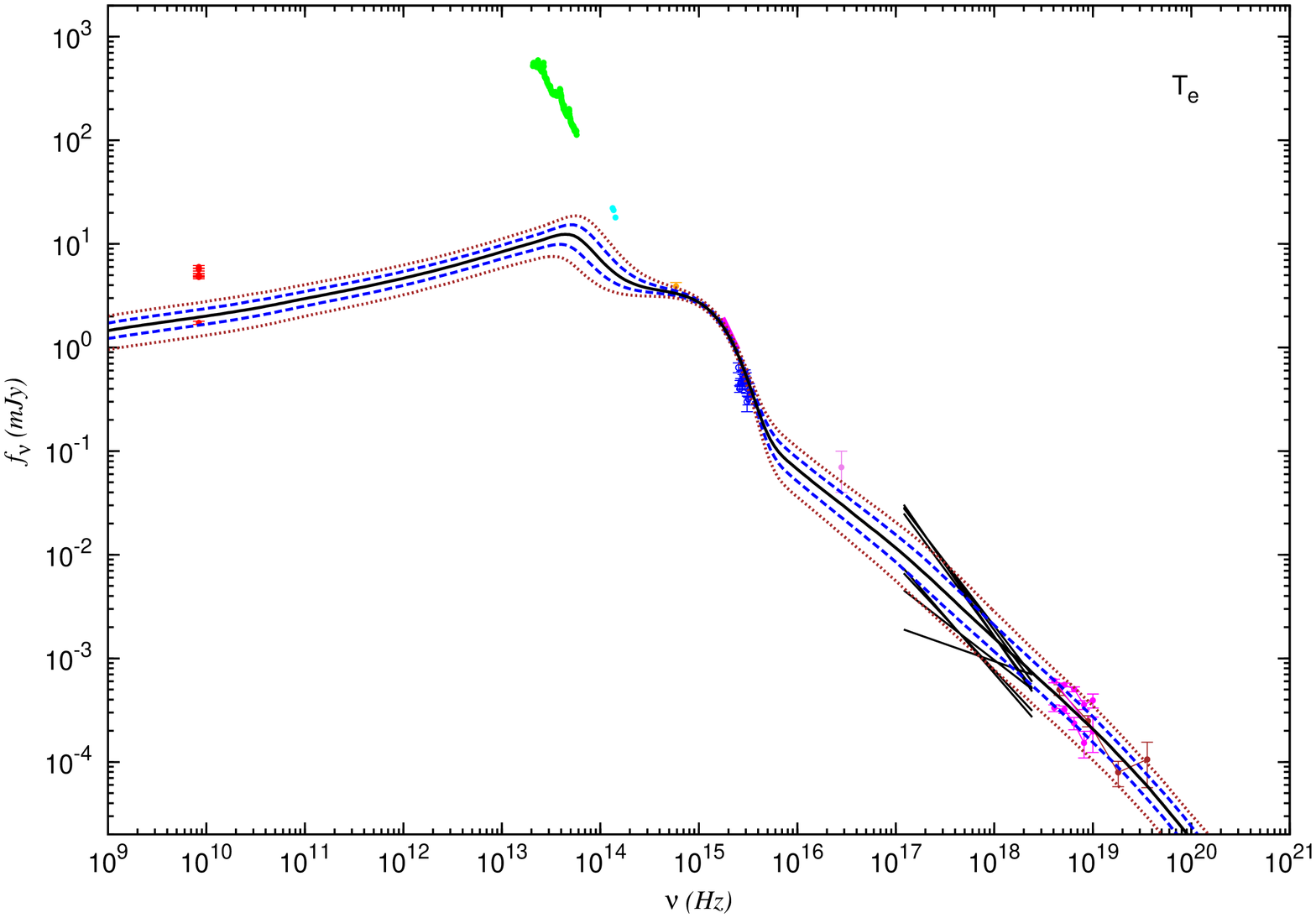} 
 \includegraphics[height=0.43\textwidth, angle=-90]{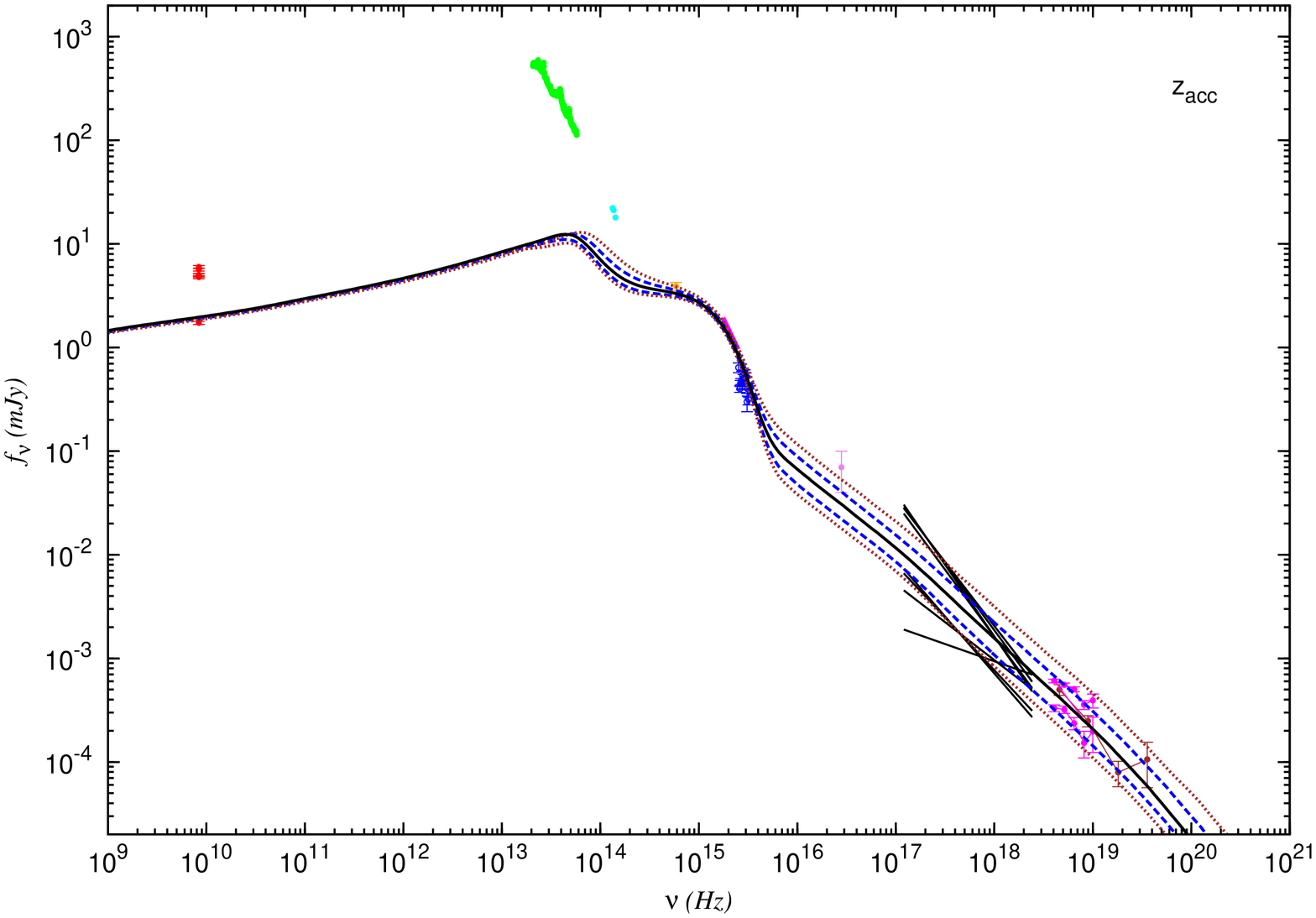} 
 \includegraphics[height=0.43\textwidth, angle=-90]{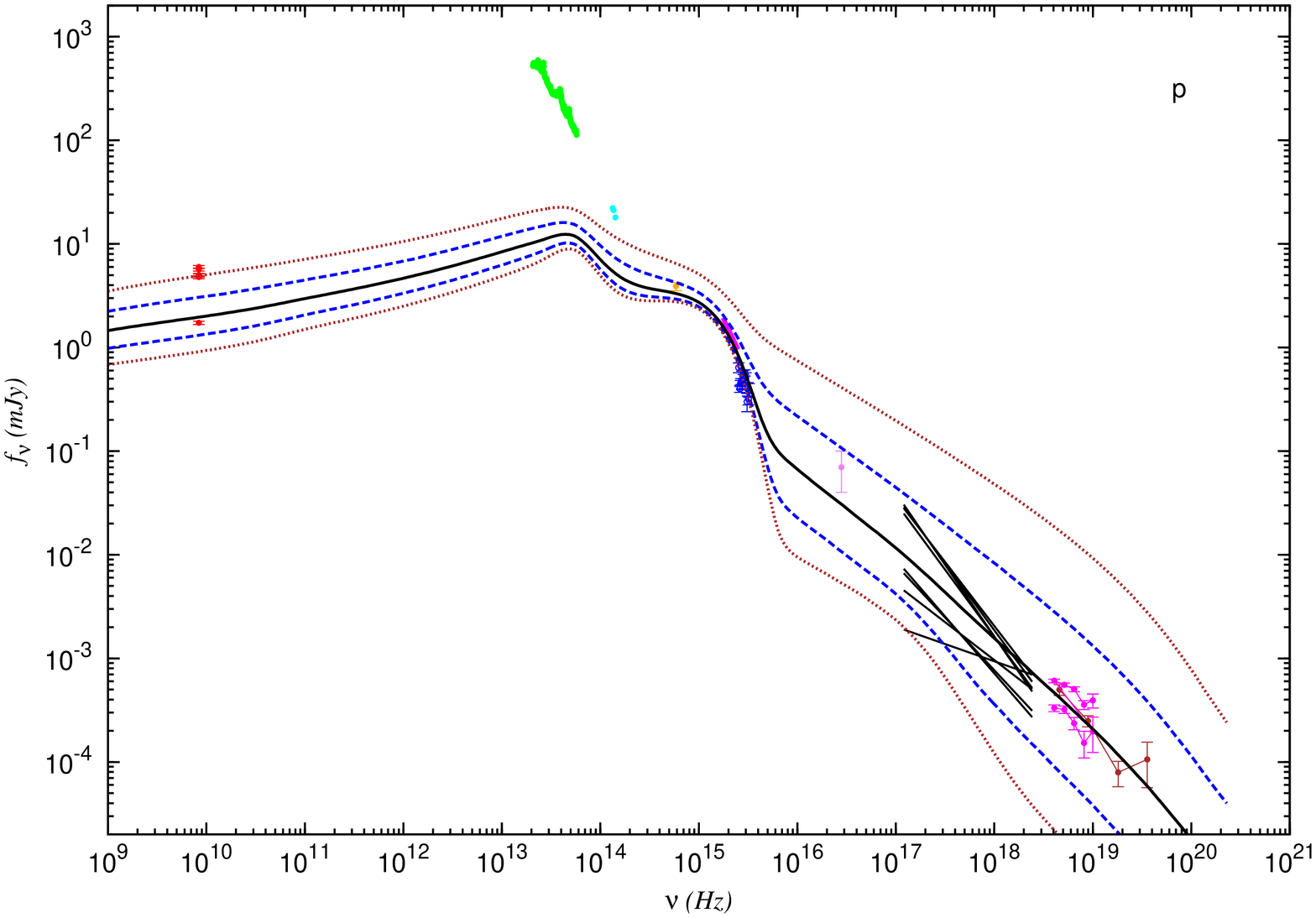} 
 \includegraphics[height=0.43\textwidth, angle=-90]{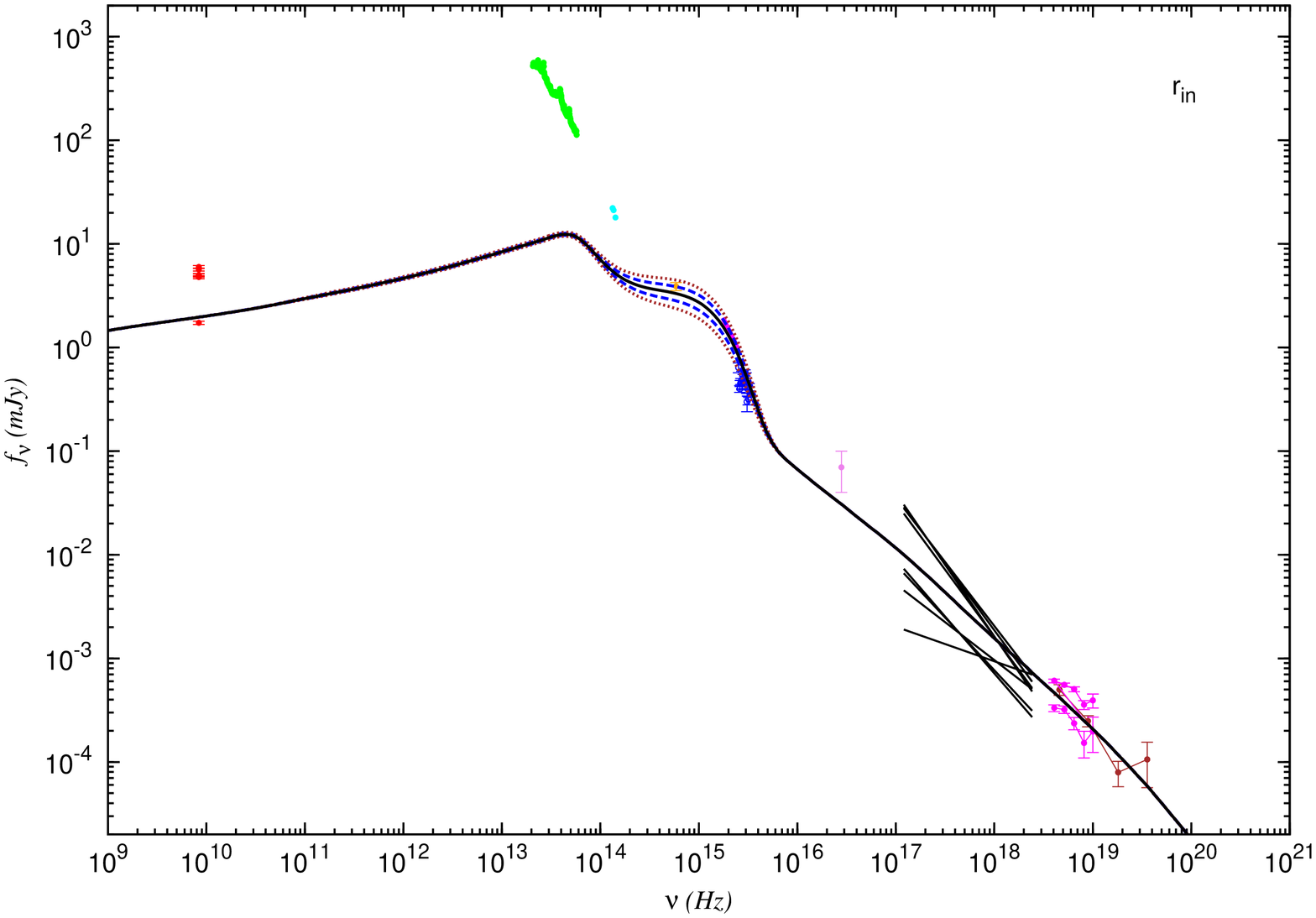} 
 \includegraphics[height=0.43\textwidth, angle=-90]{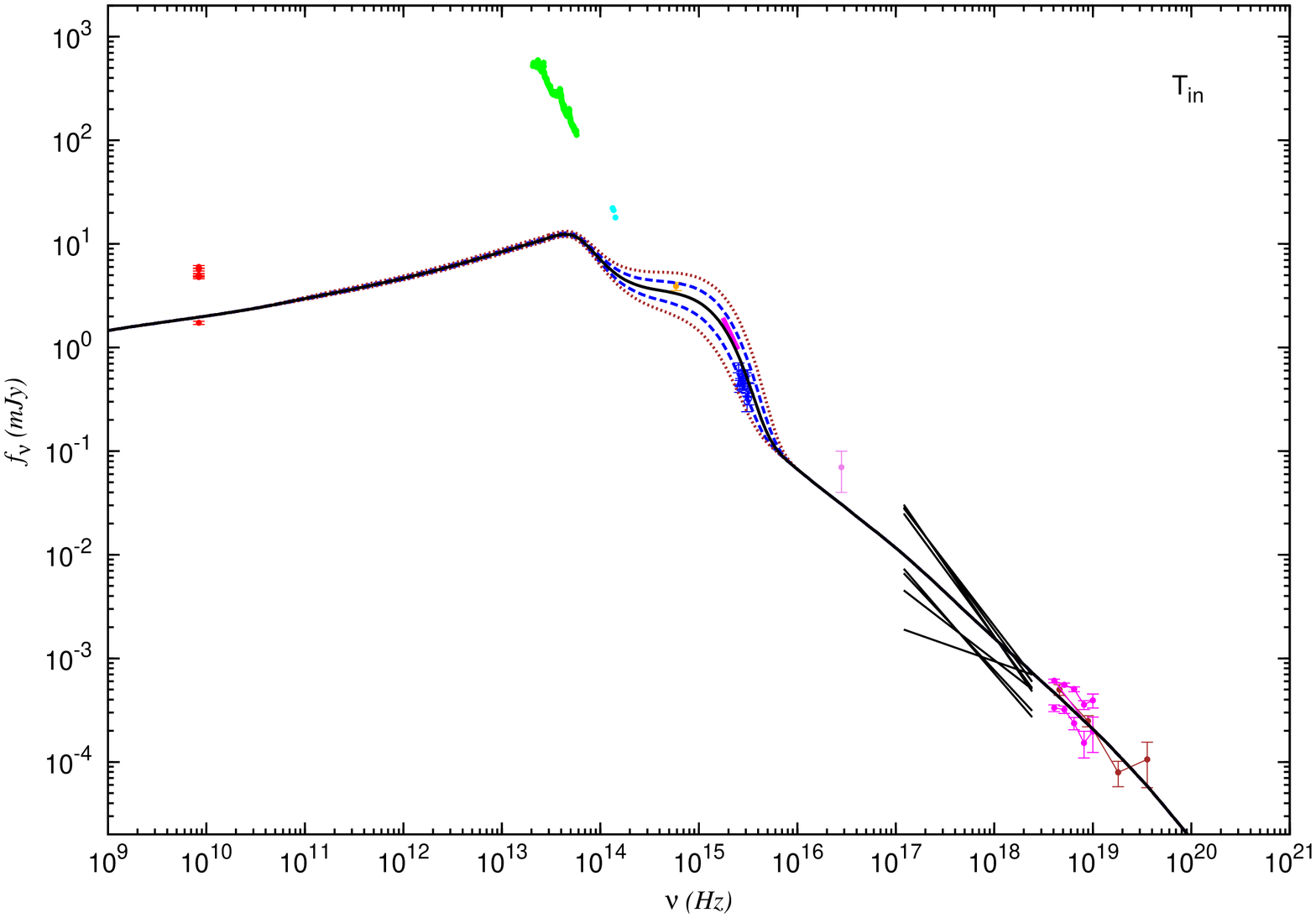} 
   \caption{%
   The solid black line in each of the above panels is the best-match
   model SED shown in Fig.~\ref{f:sed}.  The dashed blue lines show the
   SED when one of the jet model parameters (labeled near the
   top-right for each panel) was varied by $\pm$10\%. The remaining
   parameters were fixed to the best-match model.  The dotted brown lines
   show the corresponding variability in the SEDs when the model
   parameter is changed by $\pm$20\%.
   }
 \label{f:sed_pars} 
\end{figure}

Given the various uncertainties involved, e.g., in our understanding
of the cause of the large X-ray variability, non-simultaneity of
data from different wavelengths (except for the VLA/EVLA data and
the {\em Chandra} data, which were obtained simultaneously),
contribution to the optical/UV flux from non-AGN components, dust
heating, etc., a detailed statistical comparison between the data
and the model would yield errors which will significantly underestimate
the systematic uncertainties.  In other words, the uncertaintites
involved are much larger than simply the deviation of model prediction
from the data.  Therefore the best-match model here is `$\chi$-by-eye',
and is shown by the line labeled `Total' in Fig.~\ref{f:sed}.  In
Fig.~\ref{f:sed_pars} we show the model variability when one of the
parameters was varied by $\pm$10\% and $\pm$20\% from the best-match
model while the remaining parameter values were fixed to those of
the best-match model.

We adopted a distance of 10 Mpc \citep{Sani+2010} and the recently
revised reverberation mass estimate of 1.7$\times$10$^{6}$ \msun\
\citep{Denney+2009} for \src.  The outer radius of the accretion
disc is not constrained by the data and was arbitrarily set to a
value of $10^8$ \rg.  The remaining parameters were obtained from
the best-match to the data.

Assuming that the radio originates in an optically thick, stratified
jet, a comparison of the radio and \spitzer/IRS fluxes immediately
shows that the \spitzer\ fluxes are at least an order of magnitude
higher than the expected jet flux.  The large slit-width of the IRS
(3.6\arcsec) is primarily responsible for this. The high
spatial-resolution Gemini/NIFS spectrum, shows that the AGN
contribution is much less.

The soft X-ray continuum, obtained with \chandra\ simultaneously
with the 8.4 GHz \vla/{\em EVLA} observations \citep{King+2011},
shows much more variability both in the power law index and
normalization.  Such variations could arise due to rapid changes
in the particle energisation/acceleration/cooling processes occurring
in small ``knots'' in the post-shock region of the jet, e.g., as
seen in the blazar BL Lacertae by \citet{Marscher+2008}.  A full
understanding of particle acceleration in such jets is still lacking;
here we use a particle energy distribution index $p$=\valpe\ which
matches the observed $>$10$^{16}$ Hz SED quite well if the observed
EUV and X-ray emission is largely due to optically thin synchrotron.
The jet inclination was taken to be \valinclin. While not strongly
constrained by the data, a much lower value ($\lesssim$40\arcdeg)
of the inclination would cause a larger fraction of the Doppler-boosted
synchrotron photons from the outer parts of the jet to overestimate
the radio flux.  Inclinations greater than \about60\arcdeg are
similarly not preferred. We note that this inclination is same as
the results by \citet{Christopoulou+1997}, who modeled the narrow-line
region of \src\ as an outflowing biconical region of inclination
50\arcdeg, thus suggesting that the relativistic jet as well as the
much slower outflowing narrow-line region are co-aligned.  Since
we assume a symmetric model where the jets are perpendicular to the
accretion disc, this means that the inclination to the disc normal
is also \valinclin.

Since the radio data do not show any hint of low-energy cutoff, the
minimum length of the jet was set to 0.3 parsec, which is the upper
limit for the source radius of the central component detected by
\citet{GP2009} using VLBI.  Also, while searching for the best-match
model more weight was given to the VLA A configuration data point
of \citet{King+2011} since that observation had the highest spatial
resolution as well as the smallest error bar of all the radio
observations presented in that paper.

The equipartition factor ($k$) is not strongly constrained by the
data, although we could not obtain good matches either for $k<0.5$
or for $k>10$, pointing toward a near-equipartition model.  We
therefore set $k$=1. In addition to the jet, an accretion disk with
\rin=\valrin\ \rg\ and \tin=\valtin\ K is required to model the
SED.  This however assumes that the entire optical and UV emission
is from the accretion disc.  As discussed in \S\ref{s:conclusion}
(and perhaps as evidenced from the PAH lines in the \spitzer/IRS
spectrum), some of the UV emission could be from an actively
star-forming nuclear star cluster.  Some of the optical/UV radiation
could also be due to reprocessing of the X-rays from the jet base
by the outer accretion disk, as suggested by the time-lags reported
in \citet{Breedt+2010}.  Therefore the inner disc radius obtained
here is more realistically an upper limit on the actual radius.

The pre-shock synchrotron is the emission from the base/launching
region of the jet.  While the data does not constrain it directly
in this case (see, e.g., the solid orange curve in Fig.~\ref{f:sed}),
the physical conditions at the base determine the nature of the
flow farther downstream along the jet in the model.  In order to
produce the observed radio emission (which arises in the optically
thick, post-shock synchrotron regime) and the X-rays (which in our
model are originating mainly in a region close to the base where
the a fraction of the thermal leptons are accelerated to a non-thermal
powerlaw distribution), only certain physical conditions and
geometries of the base are allowed, and these choices (\ro=\valro\
\rg, \ho/\ro=\valhratio, \te=\valte\ K, \zacc=\valzacc\ \rg) lead
to the observed pre-shock synchrotron emission.  An input jet power
of \valnj\ \ledd, initially input into the particles and magnetic
field at the base of the jets, is needed to match the broadband
SED.  The inverse Compton component (solid blue line in Fig.~\ref{f:sed})
is also not strongly required by the data, however it is computed
self-consistently by the model and we include it in the plot to
show that the X-ray emission is indeed dominated by synchrotron
radiation and not by inverse Comptonization.  We also tried models
where the high energy is primarily dominated by inverse Compton
scattering.  However, within the framework of jet models, an
IC-dominated model requires very extreme and rather unphysical
parameters (e.g., \ro$<$\rg, and \ho/\ro$>$20) for the X-rays, and
is still unable to model the EUVE flux. Therefore we favor a
synchrotron origin for the EUVE and soft X-ray flux.

\section{Discussion and Conclusions} \label{s:conclusion} 
It has long been known that the so-called radio-quiet Seyfert
galaxies are not entirely radio silent (see, e.g., \citealt{UW1989},
\citealt{Wilson1991}, and also \citealt{HoPeng2001} for a discussion).
Recent VLBI observations of some of the faintest radio cores in
Seyfert galaxies have suggested the presence of jets \citep{GP2009}.
In this work we have studied the broadband nuclear emission from
\src, a narrow line Seyfert galaxy, where the presence of a jet is
suggested not only from radio VLBI, but also from observations at
other wavelengths, e.g. presence of shocked Br$\gamma$ emission
seen in high-resolution \gemini/NIFS spectrum of the nuclear region
\citep{Riffel+2008}.

The modeling presented in this work suggests that once contributions
from the host galaxy are properly accounted for, the broadband
nuclear SED of the narrow line Seyfert galaxy \src\ can be reasonably
well described by emission from an accretion disk plus a mildly
relativistic jet outflow.  A comparison of the best-match model
parameter values obtained for \src\ with XRBs in hard state and the
LLAGNs Sgr A* and M81* is given in Table~\ref{tab:models}.  The jet
parameters for \src, viz. input power (\nj), geometry of the launching
region (\ro, \ho), the temperature of the thermal plasma (\te),
equipartition factor ($k$), particle (lepton) energy distribution
index ($p$) are in the same ballpark regime for stellar mass black
hole X-ray binaries in hard state, even though the masses are
different by 5 orders of magnitude.  However, the location of the
region beyond which particles are accelerated (\zacc) appears to
be closer to the central engine for \src\ when compared to the the
LLAGNs.  Broadband SED of more NLS1s would be needed to confirm
whether \src\ is an anomalous source in this respect, or if this
is a generic feature of NLS1s.  From our modeling it appears that
the jet axis may be co-aligned with the outflowing narrow-line
region \citep[see, e.g.][]{Christopoulou+1997}.

Even though there is not enough high-quality high energy ($>$20
keV) data to constrain the efficiency of particle acceleration in
the post-shock jet, the \swift/BAT data suggests that the high
energy cutoff, if at all present, is at $\gtrsim$100 keV.  Similarly
it is not possible to strongly constrain the equipartition factor
in this case, but models near equipartition or slightly magnetically
dominated models are preferred.
The location of the region where acceleration starts (\zacc\about\valzacc\
\rg\ in this case) is still a free parameter in the model.  Recent
calculations by \citet{Polko+2010} have indicated that it is possible
to self-consistently derive the location of this acceleration region
within the self-similar magnetohydrodynamical approximation proposed
by \citet{VK2003}.  Once gravity is incorporated into this new
calculation, we will be able to reduce the number of free parameters
significantly in future versions of the jet model (Polko et al.,
in prep.).

\subsection{Variability} 
The high variability of the X-rays and the extreme-UV, and the fact
that they are simultaneous to within 1 ksec \citep{Uttley+2000}
suggests a common origin for these two bandpasses.  The variability
of the optical and lower frequencies is much smaller.  In the
framework of our model, the EUV and X-rays originate in a compact
jet close to the central \bh.  The observed high variability is
expected both because of the small size of the emission region and
relativistic beaming due to bulk motion of the jet plasma.  The
optical, on the other hand, originates from a much larger region
in the accretion disc, and therefore exhibits much lower variability.

The energy distribution index of the shocked leptons ($p=\valpe$)
obtained here is somewhat steeper than the values predicted e.g.
by diffusive shock process \citep{HeavensDrury1988}.  It is however
possible that for \src\ the EUVE and X-ray bandpasses are at a
higher energy than the cooling break energy, where the particle
energy distribution steepens from $p$ to $p+1$ due to enhanced
cooling.  If the cooling break is indeed at a frequency lower than
the EUVE, an immediate effect of this would be to move the
optically-thick-to-optically-thin transition, currently in the
\about 10--15 micron range, to submm wavelengths (where this break
has been seen for the LLAGNS Sgr A$^*$ [\citealt{FalckeMarkoff2000,
MeliaFalcke2001,Maitra+2009b}], and M81$^*$ [\citealt{Markoff+2008}]).
Since the location of the cooling break is dependent on time as
well as particle injection/acceleration rate, some of the observed
variability could also be attributed to cooling.

\citet{Ponti+2006, Miller+2010} have noted that above $\gtrsim$5
keV there may be a tendency for the rms variability to decrease
with increasing energy.  X-ray flux in our model originates in the
optically thin synchrotron from a distribution of shocked particles
beyond \zacc.  The spatial extent of such shocks may be small, and
the X-ray variability could be due to rapid changes in particle
energisation, e.g, perhaps due to magnetic reconnection events.
While the soft X-rays ($<$10--20 keV) originate mainly due to this
``post-shock'' synchrotron, it is possible that inverse Compton
scattering (of synchrotron photons produced at the base of the jet
(SSC) as well as that of the soft photons from the accretion disc)
becomes dominant at $>$20 keV.  Assuming a reasonably constant mass
accretion rate by the disc, and a constant fraction of mass channeled
into the jets, the variability of this IC component would be small,
and would largely reflect the (more slowly occurring) changes in
mass accretion rate.

The simultaneous \chandra\ and \vla\ data presented by \citet{King+2011}
suggest that during their 2008--2009 campaign \src\ shows an
anti-correlation between X-rays and radio.  Another recent work by
\citet{Jones+2011}, using {\em RXTE} and \vla\ observations made
during 2000--2001, found a weak positive correlation between radio
and X-ray fluxes.  The origin of the anti-correlation observed by
\citet{King+2011} is still not clear; however this may be due to
the following two reasons: (a) given the uncertainties in the bulk
speed of the jet plasma, the simultaneous (as observed by us) radio
and X-ray fluxes may not be causally connected, and (b) as
multiwavelength monitoring results of Sgr A* show \citep[see,
e.g.][]{Eckart+2006, YusefZadeh+2008}, there seems to be no tight
correlation between the X-ray and radio flares.  In Sgr A* the
optically thin NIR/X-ray flares most likely trace particle
acceleration/(re)energization and cooling very close to the black
hole, which only marginally affects the optically thick radio flux
\citep{Markoff+2001}.  Changes in radio flux on the other hand are
mainly driven by adiabatic expansion of overdensities in the outflow
\citep{Maitra+2009b}.  It is also possible that \src\ is in a
transitional phase between hard and soft states, where the radio
emission is in the process of getting quenched.  While comparing
with XRBs one must however keep in mind that the observed ``quenching''
of XRB jets may not mean that the jets have actually turned off.
It is possible that XRBs jets may only appear quenched because they
have become quite weak and the sensitivity is not enough to detect
them.  The presence of a separate mode for the disk--jet coupling
(i.e., separate from the coupling which gives rise to the fundamental
plane relation) operating in \src\ during the \citet{King+2011}
observations cannot be ruled out either.

\subsection{The inner-edge radius} 
Our modeling suggests that the inner edge of the accretion disk is
\about\valrin \rg.  Given that the model luminosity is $\sim$\valLeddmodel\
\ledd, a large \rin\ is not unreasonable.  The profile of the iron
K$\alpha$ emission line near 6.4 keV \citep[see, e.g.,][]{Ponti+2006,
Miller+2010}, whose origin is generally attributed to hard X-rays
irradiating the relatively colder accretion disc \citep{ReynoldsNowak2003,
Miller2007}, is quite narrow and does not show strong relativistic
broadening, further suggesting that the accretion disc may be
truncated at a large radius.  However, by comparing the observed
variability in the reflection features seen in the X-ray data with
reflection models that include light-bending, \citet{Ponti+2006}
have suggested a much smaller inner edge radius (within a few \rg).
We would like to point out that our estimation of the disc's \rin\
assumes that the optical, UV, and FUV light are entirely coming
from the accretion disc.  However, ongoing star formation, as
evidenced by the PAH lines in the \spitzer\ spectrum, may contribute
some fraction (although most likely not a large fraction; see below)
of the light that we are currently attributing to the disc.  Any
(presumably small) contribution from star formation will reduce the
disc contribution, which in turn will likely imply a smaller
disc-truncation radius (assuming that \tin\ of the accretion disk,
which determines the peak of the optical/UV SED, is not changing).

An additional contributor to the optical and UV flux is reprocessing
of the more energetic EUV/X-rays from near the central black hole.
A recent work studying the long-term optical and X-ray light curves
of \src\ by \citet{Breedt+2010} hints at a lag of $1.2^{+1.0}_{0.3}$
days, with the optical lagging behind the X-rays, thus favoring a
reprocessing origin.  On the other hand, results by \citet{Shemmer+2003}
suggest that part of the optical leads the X-ray by \about2.4 days.
Based on their detailed reprocessing model \citet{Breedt+2010}
concluded that there seems to be no common origin for the observed
X-ray/optical variability, and that not all of the observed optical
variability can be attributed to reprocessing.  Contribution by
reprocessing would tend to lower the disk \rin.  Therefore the \rin\
deduced here should perhaps more realistically be seen as an upper
limit on the actual inner radius.

An independent measurement of the inner radius may come from a
self-consistent modeling of the soft excess in this source.  While
the true nature of the soft excess is still debated \citep[see,
e.g.,][]{Gallo2006, Middleton+2007, Fabian+2009}, a survey of 34
type 1 AGNs by \citet{Crummy+2006} and detailed spectral and timing
analysis of the NLS1 galaxy 1H0707-495 by \citet{Fabian+2009} and
\citet{Zoghbi+2010} showed that the soft excess can be well modeled
by relativistically blurred photoionized disc reflection model.
Assuming a maximally rotating Kerr black hole, disc reflection model
fits to \xmm\ data of \src\ by \citet{Crummy+2006} give an inner
disc radius of \about1.24\rg.  A more physical treatment of the
scenario, where the X-ray emission from the jet-base (which can act
as a source of irradiation) is self-consistently used to compute
disc reflection features, would help in understanding soft excess
seen in the source. If the inner disc radius is indeed close to the
innermost stable circular orbit (ISCO) as suggested by the reflection
model fits, this would require most of the flux, which is attributed
to the accretion disc in this model, to have a non-disc origin.

\subsection{Stellar contribution in the optical/UV} 
The \hst/STIS spectrum given in figure 6 of \citet{Collinge+2001}
suggests that the continuum flux between 1200--1700\AA\ is approximately
constant at F$_\lambda$$\about$2$\times$10$^{-14}$ erg cm$^{-2}$
s$^{-1}$ \AA$^{-1}$, implying an observed isotropic luminosity
L$_{\rm obs}$=1.2$\times$10$^{41}$ (D/10 Mpc)$^2$ erg s$^{-1}$.  In
order to estimate the contribution from stellar flux at these
wavelengths we used the synthetic SEDs of simple stellar populations
published by \citet{Maraston2005}.  We used a model SED with
\citet{Salpeter1955} initial mass function (IMF), Solar metallicity,
and red horizontal branch morphology (HBM).  Since we are mainly
interested in the blue/UV photons, the main factor that affects the
luminosity is age, and influence of other factors (IMF, metallicity,
HBM) are comparatively much smaller.  For the chosen set of parameters,
the \citet{Maraston2005} models give SEDs as a function of the age
of the stellar population.  Given the shape of the STIS continuum,
a relatively young age of \about0.1 Gyr is needed.  Integrating the
0.1 Gyr Maraston model between 1200--1700 \AA\ gives a model
luminosity of L$_{\rm model}$=3.25$\times$10$^{33}$ erg s$^{-1}$.
Therefore L$_{\rm obs}$/L$_{\rm model}\about3.7\times10^7$.  Since
the Maraston models are normalized to 1 \msun, this implies that
\about37 million young stars would be needed within the STIS aperture
in order to have the same luminosity.  Such a stellar concentration,
while not implausibly large, is still at the high end of what has
been observed in globular clusters and in nuclear star clusters
\citep[see, e.g.][]{Cote+2006}.

However there is a very important issue here that we have not
considered:  in reality it is well known that the stars in globular
clusters are quite old, and so are the stars in the bulge.  Nuclear
star clusters typically consist of a somewhat older stellar population.
According to the \citet{Maraston2005} models, the number of stars
needed to produce the STIS observed luminosity would have to be
increased by 4 orders of magnitude if the stellar population was 1
Gyr old, instead of the assumed 0.1 Gyr.  On the other hand, recent
results from \spitzer\ spectroscopy by \citet{Sani+2010} show strong
PAH emission at 6.2$\mu$m from narrow line Seyferts, indicating
ongoing star formation within the circumnuclear region included
inside the IRS's slit-width (3.6\arcsec). Clearly, high-resolution
MIR observations would be required to better understand ongoing
star-formation in the nuclear region of NLSs.  The variability seen
in the optical wavelengths however argues against a scenario where
stellar emission dominates the total observed optical flux.  Based
on our modeling, $<$10\% of the \spitzer\ observed flux is due to
the jet.  The jet contribution is higher in the 0.1\arcsec\ \gemini\
spectrum, but still appears to be dominated by sources other than
the accretion disk and the jet.

\bigskip
The results of modeling the broadband SED of \src, along with similar
results from modeling stellar mass black holes as well as low-luminosity
AGNs like Sagittarius A$^*$ and M81$^*$, broadly suggest that the
underlying physical mechanism driving accretion inflows and outflows
do not depend on the mass of the accretor and rather most of the
observed properties scale with mass.  Evidence for the presence of
this mass-scaling is not only seen in broadband spectral modeling
and the existence of the fundamental plane, but also in various
other recent observations which include (but are not limited to)
detection of quasi-periodic oscillations in the NLS1 galaxy
RE~J1034$+$396 \citep{Gierlinski+2008,MaitraMiller2010,MiddletonDone2010},
variability (see \citealt{McHardy2010} for a recent review),
frequency-dependent lags between energy bands \citep[][and references
therein]{ArevaloUttley2006}, iron line and other associated disc
reflection features \citep[see, e.g.,][]{ReynoldsNowak2003,Miller2007}.
It is interesting to note that recent X-ray and optical observations
of gravitationally lensed quasars RX~J1131--1231 and HE~1104--1805
suggest that the X-ray emitting region of these quasars are quite
compact \citep[$\sim$10\rg\ or smaller; see,
e.g.,][]{Chartas+2009,Dai+2010}.  As our modeling of \src\ presented
here shows, such a compact region of emission is expected if the
X-rays are produced at the base of a jet.

Typically the inner disc radius is close to that of the ISCO in
XRBs in soft state, whereas the inner-disc radius we obtain here
is much larger and more characteristic of hard state XRBs where the
disc is truncated (although the radius we obtain is most likely an
upper limit as discussed above, but otherwise it would require a
rather improbably high number of extremely young stars to produce
the observed UV flux). Based on our modeling, the continuum luminosity
of the accretion disk is $6.1\times10^{41}$ erg s$^{-1}$
(\about0.3\%\ledd) and the isotropic luminosity from the power law
component between 10$^{16}$--10$^{19.5}$ Hz is $1.33\times10^{42}$
erg s$^{-1}$ (0.6\% \ledd, about twice the disk flux).  Integrating
the broadband model flux between 10$^{9}$--10$^{19.5}$ Hz and
assuming isotropic emission gives a luminosity of $2.3\times10^{42}$
erg s$^{-1}$ or \about 1.1\% \ledd, in good agreement with the
bolometric luminosity obtained with the $L_{\rm bol}\approx9\lambda
L_\lambda$ (5100 \AA) relation.  Therefore, while the timing
properties of \src\ are similar to soft state XRBs, the SED and
luminosities are more akin to hard state XRBs.  Thus it is unclear
if the XRB state definitions can be straightforwardly carried over
to AGNs, or if perhaps \src\ is in a transitional phase between
hard and soft states.

It would be interesting to test whether outflow-dominated models
can explain the SED of other Seyfert galaxies (Maitra et al. in
prep.).  As shown in this work, isolating the AGN emission from
starburst, dust etc. is one of the biggest challenges for this
project;  and it is only in the very nearby Seyferts that we currently
have the required resolution and sensitivity to see these relatively
weak jets.  The enhanced sensitivity of the EVLA at radio frequencies,
and ALMA in millimeter and submillimeter wavelengths will soon
provide data with unprecedented sensitivity and coverage, zooming
into the heart of the central engines of the nearby AGNs to study
accretion/ejection processes in much greater detail than before.

\acknowledgments
We thank Luigi Gallo for providing the \spitzer\ spectrum of \src.
DM thanks Arunav Kundu, Marta Volonteri, and Kayhan G{\"u}ltekin
for discussions about stellar population models and quasar luminosity
functions.  Additionally, DM would like to thank John Causland for
giving him a visual glimpse of \src\ through his 24'' reflector.
SM gratefully acknowledges support from a Netherlands Organization
for Scientific Research (NWO) Vidi Fellowship and from the European
Community's Seventh Framework Programme (FP7/2007-2013) under grant
agreement number ITN 215212 ``Black Hole Universe''. JMM is grateful
for support through the Chandra Guest Observer program.  We thank
an anonymous referee for suggestions that led to improvements in
the paper.


\begin{deluxetable}{llll}
\rotate
\tabletypesize{\small}
\tablecaption{Observations.
\label{tab:observations}
}
\tablewidth{0pt}
\tablehead{
\colhead{Wavelength or Energy range} & 
\colhead{Observatory/Instrument} & 
\colhead{Observation Date(s)} & 
\colhead{Reference}
}
\startdata
3.5 cm & \vla/{\em EVLA} & 2008 Dec 31 -- 2009 Jul 31 & \citet{King+2011} \\
5--14 $\mu$m & \spitzer/IRS & 2005 Dec 11 & \citet{Sani+2010} \\
2--2.4 $\mu$m & \gemini/NIFS & Jan 2006 Jan & \citet{Riffel+2008} \\
5100 \AA & MDM/1.3m, MAGNUM/2m, & 2007 Mar 20 -- 2007 Jul 30 & \citet{Denney+2009} \\
        & CrAO/(0.7m and 2.6m), & & \\
        & UNebr/0.4m & & \\
1173--1730 \AA & \hst/STIS & 2000 Mar 24--25 & \citet{Collinge+2001} \\
900--1180 \AA & \fuse & 2002 Mar 29, 2003 Jan 18, Mar 19 
     & \citet{Kaspi+2004} \\
70--190 \AA & EUVE & 1996 May 20--28, Dec 11--14 & \citet{Uttley+2000} \\
0.5--10 keV & \chandra/ACIS & 2009 Jan 07 -- 2009 Jul 31 & \citet{King+2011} \\
15--60 keV & \suzaku/PIN & 2005 Nov 10--13, 2008 Nov 6--12 & \citet{Miller+2010} \\
15--200 keV & \swift/BAT & Time averaged since start of mission & Publicly available 
\enddata
\end{deluxetable} 

\begin{deluxetable}{lllll}
\tabletypesize{\small}
\tablecaption{Comparison of jet model parameters for different sources. The
values for \src\ are based on this work, and the others taken from a 
compilation of prior published literature by \citet{Markoff+2011}. 
An asterisk (*) next to a parameter value indicates that the parameter was 
not varied while searching for the best-match model.
\label{tab:models}
}
\tablewidth{0pt}
\tablehead{
\colhead{Parameter (unit)} & 
\colhead{\src} & 
\colhead{HS-XRBs} & 
\colhead{Sgr A*} &
\colhead{M81*}\\
}
\startdata
\nj (\ledd) & \valnj	 & 10$^{-7}-10^{-1}$ & 10$^{-8}$ & 10$^{-5}$ \\
\ro (\rg)   & \valro	 & 2--100 & 2.5 & 2.4 \\
\ho/\ro     & \valhratio & 0.4--1.7 & 0.4--2.5 & 5--15 \\
\te (K)     & \valte	 & $2-5\times 10^{10}$ & 10$^{11}$ & 10$^{11}$ \\
$k$         & 1*	 & 1--5 & $>$10 & 1.4 \\
\zacc (\rg) & \valzacc	 & 10--400 & $>$10$^4$ & 144 \\
$p$         & \valpe	 & 2.4--2.9 & $>$3.8 & 2.4 \\
\enddata
\end{deluxetable} 

\end{document}